\documentclass[final]{article}
\usepackage{url}

\usepackage{graphicx}%
\usepackage{multirow}%
\usepackage{amsmath,amssymb,amsfonts}%
\usepackage{mathrsfs}%
\usepackage{xcolor}%
\usepackage{textcomp}%
\usepackage{manyfoot}%
\usepackage{booktabs}%
\usepackage{algorithm}%
\usepackage{algorithmicx}%
\usepackage{algpseudocode}%
\usepackage{listings}%
\usepackage{subcaption}%

\begin{document}
\title{Time-series Neutron Tomography of Rhizoboxes}
\author{A.~P.~Kaestner\textsuperscript{1} \and S. Di Bert\textsuperscript{2} \and R. Jia\textsuperscript{2} \and
P. Benard\textsuperscript{2} \and A. Carminati\textsuperscript{2}\\[1ex]
{\small \textsuperscript{1} PSI Center for Neutron and Muon Sciences, Paul Scherrer Institute, Villigen, Switzerland}\\
{\small \textsuperscript{2} Department of Environmental Systems Science, ETH Zurich, Zurich, Switzerland}
}

\maketitle
\begin{abstract}
This study investigates the feasibility of tilt-series neutron tomography for analyzing rhizoboxes used in root-soil interaction studies. Traditional neutron imaging methods are limited by constrained root growth volumes and poor penetration in moist soil. Using a vertical acquisition axis, the tilt-series approach avoids constraints typical of laminography. This method allows simultaneous radiographic and tomographic data acquisition, enhancing time resolution and providing detailed insights into root networks and soil water content. Experiments involved scanning six-week-old maize plants in rhizoboxes filled with sand. Results show that tilt-series tomography can effectively reconstruct root networks despite some artifacts from the missing wedge. While the tilt-series tomographic data qualitatively reveal water distribution changes, radiographic data remain essential for quantitative analysis. This approach demonstrates the potential for dynamic root-soil interaction studies, offering a valuable tool for agricultural and environmental research by providing comprehensive insights into the rhizosphere.
\end{abstract}

\section{Introduction}\label{sec_introduction}
Processes in the rhizosphere are well-studied using neutron imaging. This task is challenging as it requires compromises in sample design and imaging conditions. Studying roots requires a certain unconstrained volume for the roots to grow unhindered on one side, while the neutrons limit the rhizosphere volume due to the short penetration depth in moist soil. This resulted in two main rhizobox designs for neutron imaging experiments. For young plants, it is possible to use round soil samples with dimensions suitable for tomography. This geometry can only be used until the root network grows to be bounded by the container walls and is unnaturally bent to become clustered at the walls. Older plants have much larger root networks and would be too constrained by the wall of the small cylinders. Instead, slabs are used for these plants. These samples are often designed for radiography time series acquisition as the thin slab can be considered an approximately two-dimensional representation of the distribution of the material. The slab is also limited as the roots can only grow unconstrained in two directions, while the remaining direction is narrowed down to a minimum. This approximation initially works well, but at some point, with refined models, it becomes increasingly important to obtain information about the location of sample features in the beam direction. This information can only be obtained using computed tomography. The slab is far from the ideal shape for tomography because of the unbalanced aspect ratio, making it impossible for the neutron beam to penetrate the sample in the direction parallel to the plane. 

Two approaches usually obtain tomographic images from slabs: a full turn with a tilted acquisition axis or incomplete sample rotation with a vertical acquisition axis \cite{helfen2011_nLamino,osterloh2011limited}. Scans with a tilted axis are often referred to as laminography. The incomplete scan lacks a wedge of the sample rotation and has different names depending on the application: tomosynthesis (medicine)\cite{Levakhina2014} and tilt series (microscopy). Both cases have to compromise because the Radon space is incompletely filled, which results in more or less severe artifacts in the reconstructed images. The tilt series were identified as less valuable due to the nature of the artifacts caused by the missing wedge in the Radon space \cite{Xu2012}, which are more severe and also asymmetric compared to laminography. We argue, however, that this method is still relevant for specific use cases where the sample is unsuited for the rotation around a tilted acquisition axis. The sample may contain liquids or loosely packed grains that could rearrange or the sample may be wired in a manner that constrains its ability to perform a complete rotation. The use of laminography for rhizoboxes was demonstrated by Rudolph-Mohr et al. \cite{RudolphMohr2021}. This approach requires further compromises for the scan; the content of the rhizobox must be fixed to avoid reordering the slab content, and it is not possible to perform a complete turn scan if the shoot is intact. Furthermore, combining radiography and tomography scans is impossible as the sample needs two different mountings. These compromises are obsolete with tilt-series as the sample will be scanned using a vertical acquisition axis.

The aim of this study is to investigate the feasibility of using a tilt series for rhizoboxes. We will explore the method's limitations and compare the results to a full-turn scan. We will also investigate the possibility of using a dynamic scan sequence to obtain radiographs and tomographic data in the same scan. This method will provide a time series of radiographs and tomographic data with flexible time resolution. The method will be demonstrated on a maize plant in a rhizobox. The results will be used to quantify the root network morphology and water content in the rhizobox. The results will be compared to a full-turn scan. The study will provide a method for obtaining quantitative data from rhizoboxes using a tilt series. The method will be useful for studying root-soil interactions in the rhizosphere. Due to the reconstruction artifacts, the missing wedge tomography can not be used to quantify the water content in the rhizobox. Radiography time series are needed for this. At first sight, this looks like two separate scans. However, we will show that the two can be interleaved in the same scan sequence. This will provide a time series of radiographs and tomographic data with flexible time resolution. The tomography time series will be achieved using the golden ratio to determine the scan angle sequence \cite{kaestner2011_golden},\cite{Craig_2023}.

\section{Methods and samples}\label{sec:methods}

\subsection{Sample design}\label{sec:sample_design}
\begin{figure}[ht!]
    \centering
    \begin{subfigure}[b]{0.4\textwidth}
        \includegraphics[width=\textwidth]{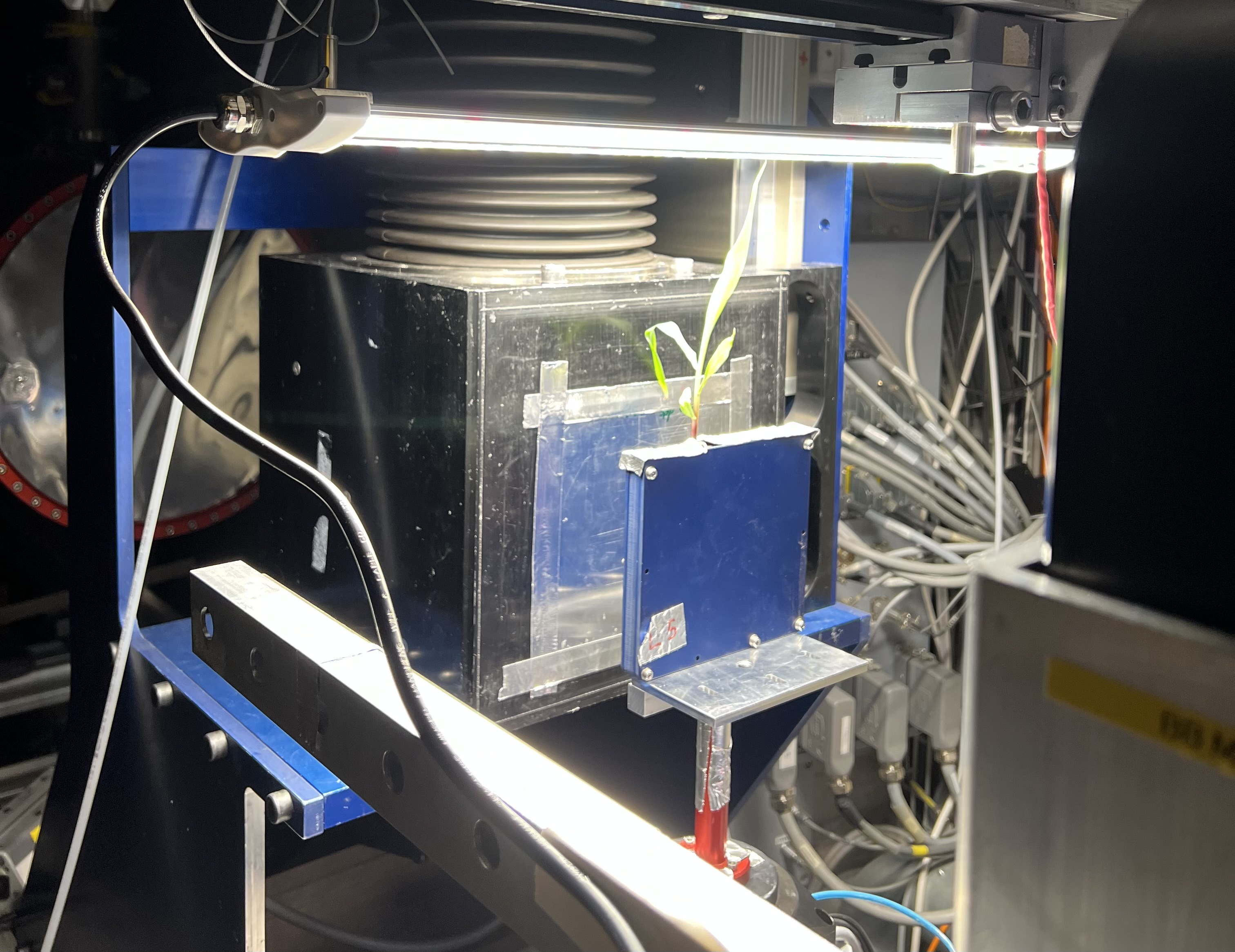}
        \caption{}
        \label{fig:ExperimentPhoto}
    \end{subfigure}
    \hfill % optional: add some horizontal spacing
    \begin{subfigure}[b]{0.5\textwidth}
        \includegraphics[width=\textwidth]{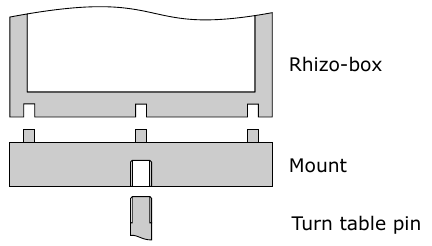}
        \caption{}
        \label{fig:RhizoBoxDrawing}
    \end{subfigure}
    \caption{A rhizo-box with a maize plant installed for scanning on the experiment position at the neutron imaging instrument ICON, Paul Scherrer Institut (\subref{fig:ExperimentPhoto}). The mounting consists of two parts, one mounted on the turn table and the rhizo-box with matching holes in the bottom plate (\subref{fig:RhizoBoxDrawing}).}
    \label{fig:rhizobox}
\end{figure}

The rhizoboxes are designed to study root-soil interactions, figure \ref{fig:ExperimentPhoto}. The rhizoboxes used in the following study are made of aluminum to minimize attenuation from the box. The box has the outer dimensions 140$\times$140$\times$16~mm\textsuperscript{3} and the inner dimensions for the soil slab 125$\times$125$\times$10~mm\textsuperscript{3}. The mounting is solved using a base mount with three pins in-line attached to the turn table, and each rhizobox has matching holes in the bottom plate, figure \ref{fig:RhizoBoxDrawing}. This mounting allows for precise repositioning of the sample in the beam. This is important as the experiments often involve several boxes that will be scanned at different times of the day or before and after treatment.

The rhizoboxes are filled with sand material with a porosity of 42\% and planted with maize seeds. The maize plants were grown for six weeks to produce a root network filling out the rhizobox.  

\subsection{Determining the tilt range}
The choice of the tilt angle for a neutron tilt series is balanced between how well sample features need to be represented on one side and the ability to penetrate the sample and blurring on the other side. The fidelity of the reconstructed features to shapes in the sample requires a full tomography scan. In contrast, the maximum scan range is limited by transmission, and the penumbra blurring is caused by the required distance between the sample and detector for a given scan wedge. 

\begin{figure}[ht!]
    \centering
    \begin{subfigure}[b]{0.3\textwidth}
        \includegraphics[width=\textwidth]{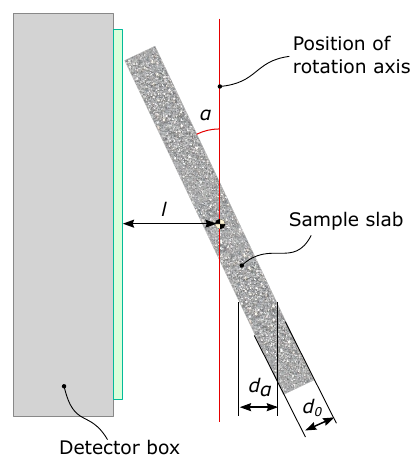}
        \caption{}
        \label{fig:tiltedslab}
    \end{subfigure}
    \hfill % optional: add some horizontal spacing
    \begin{subfigure}[b]{0.65\textwidth}
        \includegraphics[width=\textwidth]{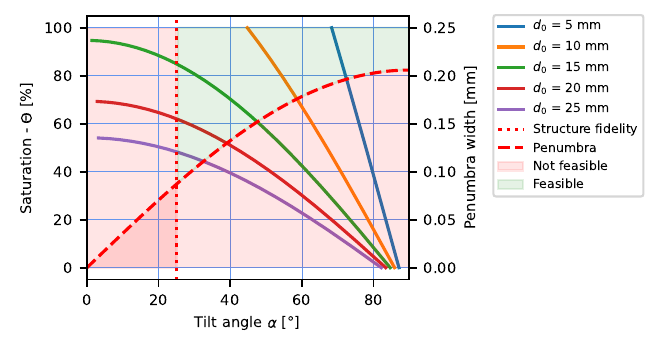}
        \caption{}
        \label{fig:limitations}
    \end{subfigure}
    \caption{The setup geometry (\subref{fig:tiltedslab}) limits the tilt angle, $\theta$. The greatest tilt angles for different sample thicknesses and saturation levels in a 42\% porosity material.}
    \label{fig:tiltlimitations}
\end{figure}

The transmission is primarily determined by the porosity of the soil material, the slab thickness, and water content. Beer-Lambert's law gives the transmission at oblique beam incidence to the slab plane. In the case of a rhizobox, this equation reduces to a sum of three materials: water, sand, and the wall of the sample container. The container is made of aluminum. Thus, we can write the transmission as
\begin{equation}
    T = \frac{I}{I_0} = e^{-\left(\mu_{\text{Al}}\cdot d_{\text{Al}} + \mu_{\text{Sand}}\cdot d_{\text{Sand}} + \mu_{\text{H2O}}\cdot d_{\text{H2O}} \right)},
    \label{eq:transmission_model1}
\end{equation}
The wall and sand thicknesses are constant, while the water length depends on the sample water saturation, $\Theta$. The length through the sample at different view angles increases with the tilt angle, $\alpha$. Equation \ref{eq:transmission_model1} can be revised into
\begin{equation}
    T = \frac{I}{I_0} = e^{-\left(\mu_{\text{Al}}\cdot d_{\text{Al}}+ \mu_{\text{Sand}}\cdot d_{\alpha} \cdot (1-\phi_{Sand}) + \mu_{\text{H2O}}\cdot d_{\alpha} \cdot \Theta \cdot \phi_{Sand} \right)},
    \label{eq:transmission_model2}
\end{equation}
when we take the thickness $d_{\alpha}=d_{0}/\cos\alpha$, porosity $\phi_{Sand}$, and saturation $\Theta$ of the pores into account. The thickness and saturation are variable during the experiment. We aim for a transmission greater than 10\% to calculate the maximum tilt angle. Lower transmission is not advisable as noise and scattering amplitudes are in this order of magnitude. In figure \ref{fig:limitations}, we show the greatest allowed tilt angle for a sand material with $\phi_{sand}$=42\% for the rhizobox described above in section \ref{sec:sample_design}. The greatest tilt angle depends on the degree of water saturation and slab thickness. Here, we used $d_0$=10~mm. The red band to the left indicates tilt ranges that produce too strong missing wedge artifacts. The tilt series scan will be in the interval $\alpha_{scan}=[-\alpha, \alpha]$. The tilt angle further impacts the distance between the sample and detector, $l$ in figure \ref{fig:tiltedslab}. This distance causes a penumbra blurring given by the collimation ratio of the neutron beam. The blurring represented by the vertical axis to the right in figure \ref{fig:limitations} is computed for the rotation axis. In reality, $l$ varies between just a few mm to $2 l$ during the scan. In this plot, we used L/D=340, corresponding to the ICON's nominal value when the 20~mm neutron aperture is used \cite{kaestner2011_ICON}.  

The distance between the sample and the detector determines the width of the penumbra blurring, which is the major source of unsharpness in the setup used. 
A lower bound on the tilt angle is given by the severity of missing wedge artifacts. These include fidelity to the original shape of objects in the sample and the amount of cross-talk in the beam direction. The artifacts are more severe for narrow scan ranges.  

\subsection{Dynamic scan}
Typical rhizobox experiments are performed as a time series of radiographs. A method that has proven to be useful for the study of root-soil interactions. The time series is used to quantify the dynamics of water movement in the soil and root network. The radiographs have a limitation in that they are two-dimensional and do not provide information about the location of features in the beam direction. This information can only be obtained using tomography. The dynamics of the sample require the use of golden ratio sequences \cite{kaestner2011_golden}, which have two important features: 1, motion artifacts are minimized, and 2, any sequence of consecutive projections in time covers the full range of tilt angles. The second feature is important as it allows the scan to be broken down into a series of sub-scans, each covering the full scan range with an almost uniform distribution of the acquisition angles. Craig \emph{et al.} \cite{Craig_2023} showed that the golden ratio sequence works equally well for tilt series as for the complete scan. Here, we modify the original golden ratio sequence to only cover a limited scan range of $\pm \alpha_{max}$. The acquisition angles are, thus, given by
\begin{equation}
    Golden(i) = \left(i\cdot 2\alpha_{max} \cdot \phi \mod 2\alpha_{max}\right)-\alpha_{max},
\end{equation}
where $\phi=(1+\sqrt{5})/2$ is the golden ratio, and $\alpha_0$ is the greatest tilt angle from the view where the slab is parallel to the detector. This sequence can be used for $i\rightarrow\infty$ or until the experiment is terminated without repeating any angles, thanks to the irrational nature of the golden ratio.

In our experiment, we are interested in obtaining both radiographs for their capacity to quantify the local water content and tomographic data for the root network topology. Normally, the radiography time series and tomography would be acquired separately after each other. This would, however, mean that the tomography can only capture the observed process's initial and final state. Thus, we propose to interleave the radiographs with the tomographic data in the same scan sequence %, figure \ref{fig:scan_sequence}, 
such that,
\begin{equation}
    \begin{cases}
    i=2n & \rightarrow \alpha=Golden(i/2)\\
    i=2n+1 & \rightarrow \alpha=0^{\circ}.
    \end{cases}
    \label{eq:flipflapscan}
\end{equation}
This scan provides a time series of radiographs and tomographic data with flexible time resolution, observing the same process. The tomographic temporal resolution is naturally lower due to the number of projections needed for the reconstruction.

\section{Experiments and analysis}
\subsection{Tomography scans}
The tomography data was acquired at the ICON neutron imaging beamline at PSI \cite{kaestner2011_ICON} using the so-called midi setup on experiment position 2. The setup was configured with an Andor iKon L camera, an Otus 55mm lens and a 30~$\mu$m thick GadOx scintillator. This combination resulted in a FOV of 150$\times$150mm\textsuperscript{2} and the pixel size 90.9~$\mu$m.
\paragraph{Interleaved time-series scan}
The rhizoboxes were scanned in a day-night-day cycle, and plant lights were installed above the plants during the scan. The plants are expected to consume more water during the day than at night, which will show in soil water distribution. 

The demonstrated scan was obtained during the daytime and consisted of 320 radiographs. From which every second is a radiograph at 0$^{\circ}$, and the other is obtained from the golden ratio sequence in the interval $\alpha=\pm \mathrm{60}^{\circ}$.

\paragraph{Full-turn scan}
At the end of the time series, a full 360$^{\circ}$ tomography scan was performed to provide a reference. The slab is assumed to be sufficiently dry to allow reconstruction. The slab was moved further from the detector to allow this scan without colliding with the equipment.

\subsection{Processing projection data}
The processing has two tracks, one for the radiography data and the other for the tomography.
Both data sets were normalized using the scattering correction method described in \cite{Boillat2018_BB} and \cite{CarminatiBB_2019} to reduce artifacts originating from scattered neutrons. The data is further also filtered to remove outlier spots and ring artifacts. 

The radiography data was processed using Python scripts based on NumPy and SciKit-Image packages.

The tomography data was reconstructed using the filtered back projection algorithm provided in the tomography reconstruction tool MuhRec \cite{kaestner2011_muhrec}, \cite{kaestner2023_muhrec442} includes support for golden ratio projection data and weighting needed to handle the missing wedge. All pre-processing was done in MuhRec.

\section{Results}\label{sec_results}
\subsection{Tilt range}
\begin{figure*}[ht!]
    \centering
    \includegraphics[width=\linewidth]{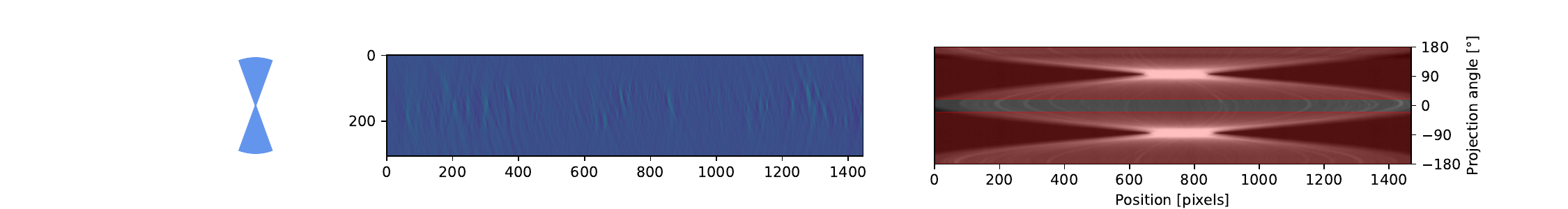}
    \includegraphics[width=\linewidth]{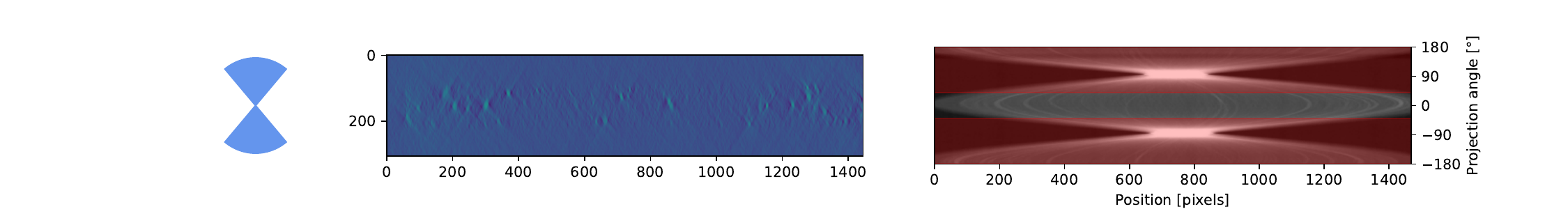}
    \includegraphics[width=\linewidth]{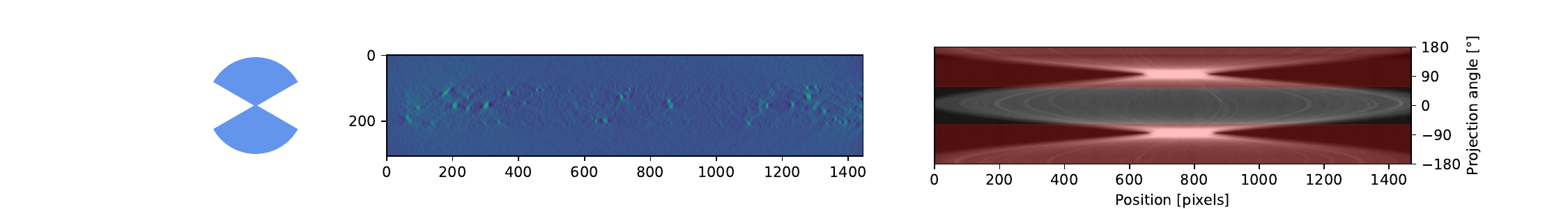}
    \includegraphics[width=\linewidth]{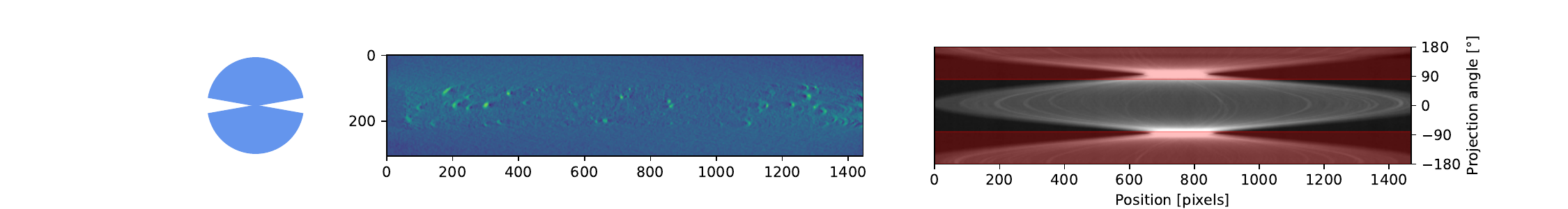}
    \caption{Reconstructions from the full scan with covered wedges from 20$^{\circ}$ and 80$^{\circ}$}. The sinograms to the right show each slice's used fraction of the full sinogram.
    \label{fig:tiltrange}
\end{figure*}

The full tomography of a dry slab allowed us to reconstruct the entire root network. This scan was reduced to several sub-scans with wedge sizes between 20$^{\circ}$ and 80$^{\circ}$. The reconstructed slices are shown in figure \ref{fig:tiltrange}. The reconstructions clearly show the degradation of sample features when the covered wedge size decreases. The choice of using $\alpha=\pm \mathrm{60}^{\circ}$ is confirmed. The missing wedge artifacts are present, but the main shape has still been well reconstructed. 

\subsection{Time-series}
\begin{figure*}[ht!]
    \centering
    \includegraphics[width=\linewidth]{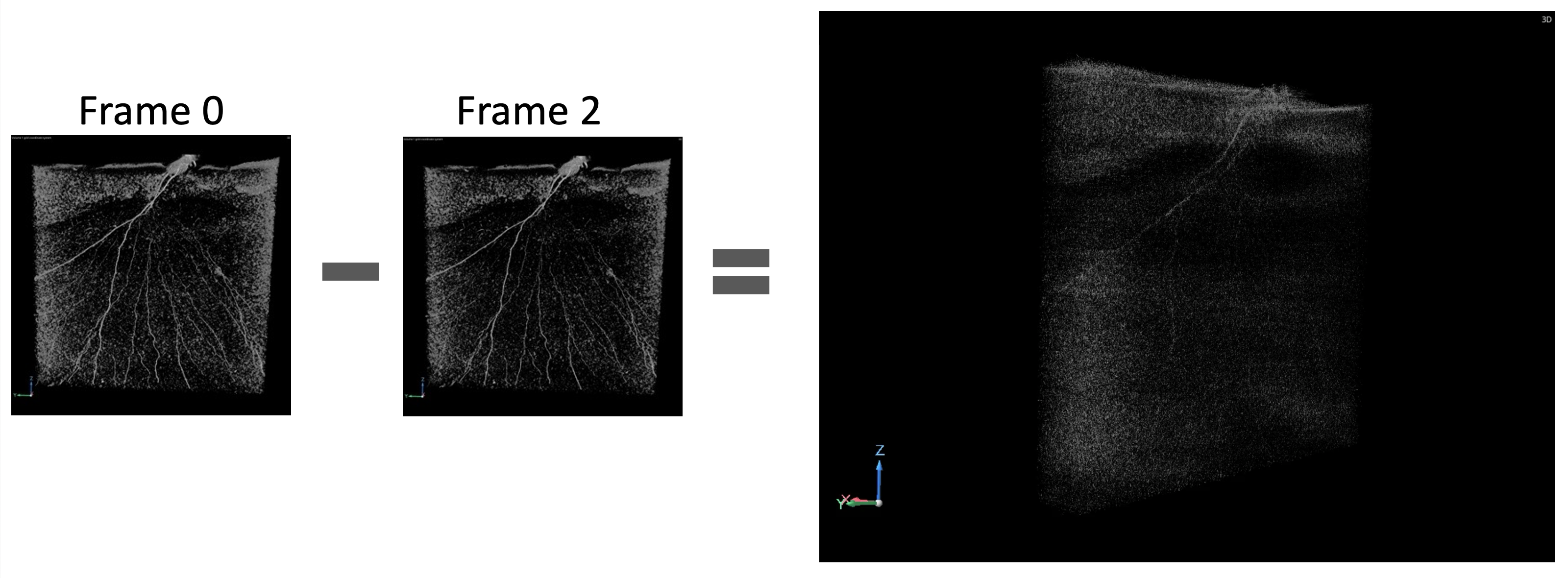}
    \caption{The difference between the last and first volume time frame in the scan sequence.}
    \label{fig:timedifference}
\end{figure*}

The time series scan consists of 160 golden ratio projections, which we reconstructed into three time frame volumes. Each volume does not show much change at first sight, but subtracting frame 0 from frame 2 shows the change in water contents, fig. \ref{fig:timedifference}. 
Due to bias and reconstruction artifacts, the missing wedge volumes cannot be used for quantitative water content analysis. It is, however, still possible to qualitatively identify the location of the water from the tomography.

\section{Conclusions}
These preliminary results using missing wedge tomography combined with radiography for time-series acquisition of water movement in rhizoboxes show a promising path towards improving model accuracy for soil-root interactions. Upcoming experiments will focus on obtaining longer time series and systematic studies to improve image quality. Iterative reconstruction techniques are also expected to improve the image quality.

\subsection*{Acknowledgement}
This work is based on experiments performed at the Swiss spallation neutron source SINQ, Paul Scherrer Institute, Villigen, Switzerland.
\bibliographystyle{plain}
\bibliography{slabbibliography}% common bib file

\begin{thebibliography}{10}

\bibitem{Boillat2018_BB}
P.~Boillat, C.~Carminati, F.~Schmid, C.~Gr{\"u}nzweig, J.~Hovind, A.~Kaestner,
  D.~Mannes, M.~Morgano, M.~Siegwart, P.~Trtik, P.~Vontobel, and E.H. Lehmann.
\newblock Chasing quantitative biases in neutron imaging with
  scintillator-camera detectors: a practical method with black body grids.
\newblock {\em Optics Express}, 26(12):15769, jun 2018.

\bibitem{CarminatiBB_2019}
C.~Carminati, P.~Boillat, F.~Schmid, P.~Vontobel, J.~Hovind, M.~Morgano,
  M.~Raventos, M.~Siegwart, D.~Mannes, C.~Gruenzweig, P.~Trtik, E.~Lehmann,
  M.~Strobl, and A.~Kaestner.
\newblock Implementation and assessment of the black body bias correction in
  quantitative neutron imaging.
\newblock {\em {PLOS} {ONE}}, 14(1):e0210300, jan 2019.

\bibitem{Craig_2023}
Timothy~M. Craig, Ajinkya~A. Kadu, Kees~Joost Batenburg, and Sara Bals.
\newblock Real-time tilt undersampling optimization during electron tomography
  of beam sensitive samples using golden ratio scanning and {RECAST}3d.
\newblock 15(11):5391--5402.

\bibitem{helfen2011_nLamino}
L.~Helfen, F.~Xu, B.~Schillinger, E.~Calzada, I.~Zanette, T.~Weitkamp, and
  T.~Baumbach.
\newblock Neutron laminography-a novel approach to three-dimensional imaging of
  flat objects with neutrons.
\newblock {\em Nuclear Instruments and Methods in Physics Research A},
  651({1}):135--139, SEP 21 2011.

\bibitem{kaestner2023_muhrec442}
A.~Kaestner, C.~Carminati, D.~Tasev, and W.~Potrzebowski.
\newblock Muhrec ver. 4.2.2, December 2023.

\bibitem{kaestner2011_muhrec}
A.P. Kaestner.
\newblock {MuhRec} -- a new tomography reconstructor.
\newblock {\em Nuclear Instruments and Methods in Physics Research A},
  651(1):156--160, September 2011.

\bibitem{kaestner2011_ICON}
A.P. Kaestner, S.~Hartmann, G.~K\"{u}hne, G.~Frei, C.~Gru\"{u}zweig, L.~Josic,
  F.~Schmid, and E.H. Lehmann.
\newblock The {ICON} beamline - {A} facility for cold neutron imaging at
  {SINQ}.
\newblock {\em Nuclear Instruments and Methods in Physics Research A},
  659(1):387 -- 393, 2011.

\bibitem{kaestner2011_golden}
A.P. Kaestner, B.~M\"{u}nch, P.~Trtik, and L.G. Butler.
\newblock Spatio-temporal computed tomography of dynamic processes.
\newblock {\em Optical Engineering}, 50(12):123201, 2011.

\bibitem{Levakhina2014}
Yulia Levakhina.
\newblock {\em Three-Dimensional Digital Tomosynthesis}.
\newblock Springer Fachmedien Wiesbaden, 2014.

\bibitem{osterloh2011limited}
Kurt Osterloh, Daniel Fratzscher, Mirko Jechow, Thomas B{\"u}cherl, Burkhard
  Schillinger, Andreas Hasenstab, Uwe Zscherpel, and Uwe Ewert.
\newblock {\em Limited view tomography of wood with fast and thermal neutrons},
  volume 128.
\newblock Deutsche Gesellschaft f{\"u}r Zerst{\"o}rungsfreie Pr{\"u}fung
  (DGZfP), 2011.

\bibitem{RudolphMohr2021}
Nicole Rudolph-Mohr, Sarah Bereswill, Christian Tötzke, Nikolay Kardjilov, and
  Sascha~E. Oswald.
\newblock Neutron computed laminography yields 3d root system architecture and
  complements investigations of spatiotemporal rhizosphere patterns.
\newblock {\em Plant and Soil}, 469(1-2):489--501, 2021.

\bibitem{Xu2012}
Feng Xu, Lukas Helfen, Tilo Baumbach, and Heikki Suhonen.
\newblock Comparison of image quality in computed laminography and tomography.
\newblock 20(2):794.

\end{thebibliography}
\end{document}